\documentclass[twocolumn,aps,pre,showpacs,floatfix]{revtex4}
\usepackage{graphicx}
\usepackage{amssymb}
\usepackage{amsmath}
\usepackage{bm}
\begin{document}
\title{Chaotic synchronizations of spatially extended systems \\
as non-equilibrium phase transitions}
\author{M. Cencini}
\affiliation{INFM-CNR, SMC Dipartimento di Fisica Universit\`a Roma 1, P.zzle
A. Moro 2, 00185 Roma, Italy} 
\affiliation{Istituto dei Sistemi Complessi - CNR, via dei Taurini 19, 00185 Roma, Italy}
\author{C.J. Tessone}
\affiliation{Chair of Systems Design, ETH Z\"urich, Kreuzplatz 5, 8037
Z\"urich, Switzerland}
\author{A. Torcini}
\affiliation{Istituto dei Sistemi Complessi -  CNR, via Madonna del Piano, 10, 50019 Sesto Fiorentino, Italy}
\affiliation{INFN - Sezione di Firenze and CSDC, via Sansone 1, 50019 Sesto Fiorentino, Italy}
\affiliation{Centre de Physique Th{\'e}orique, Campus de Luminy, 13288 Marseille, France}
\date{\today}
\begin{abstract}
Two replicas of spatially extended chaotic systems synchronize to a
common spatio-temporal chaotic state when coupled above a critical strength.
As a prototype of each single spatio-temporal chaotic system 
a lattice of maps interacting via power-law coupling is considered.
The synchronization transition is studied as a non-equilibrium phase
transition, and its critical properties are analyzed at varying the
spatial interaction range as well as the nonlinearity of the dynamical
units composing each system. In particular, continuous and
discontinuous local maps are considered. In both cases the transitions
are of the second order with critical indexes varying with the
exponent characterizing the interaction range. For discontinuous maps
it is numerically shown that the transition belongs to the {\it
anomalous directed percolation} (ADP) family of universality classes,
previously identified for L{\'e}vy-flight spreading of epidemic
processes. For continuous maps, the critical exponents are different
from those characterizing ADP, but apart from the nearest-neighbor
case, the identification of the corresponding universality classes
remains an open problem. Finally, to test the influence of
deterministic correlations for the studied synchronization
transitions, the chaotic dynamical evolutions are substituted by
suitable stochastic models. In this framework and for the discontinuous
case, it is possible to derive an effective Langevin description that
corresponds to that proposed for ADP.
\end{abstract}
 \maketitle

{\bf 

Synchronization is ubiquitous in Nature: neuronal populations, cardiac
pacemakers, Josephson circuits and even coupled chaotic systems can
synchronize during their activity. Remarkably, all these different
phenomena can be described within a common framework represented by
nonlinear dynamics~\cite{PikoBook}.

Synchronization of spatially extended chaotic systems is particularly
interesting as it allows for transferring concepts and methods
borrowed from statistical mechanics to nonlinear dynamics. Indeed,
a direct connection between the critical properties of
non-equilibrium phase transitions and synchronization processes has
been established~\cite{baroni,PA02}. Noticeably,
for diffusively coupled chaotic systems only two universality classes 
-- Directed Percolation and Multiplicative Noise -- encompass 
all the synchronization transitions. Originally these classes
have been identified in completely different contexts such 
as epidemics spreading and pinning/depinning of interfaces to/from a
substrate~\cite{haye,munoz_review}. The parallel between
non-equilibrium critical phenomena and synchronization processes
is possible only thanks to the erratic nature of the synchronized 
state, where chaos somehow mimics the presence of thermal noise in 
real systems.

In this paper we propose an extension of such an analogy to
synchronization transitions of long-range coupled systems. Long range
interactions naturally appears in many circumstances ranging from
neuronal networks to solid state physics. Therefore, a
characterization of these transitions is definitely worth and of
interest for a wide scientific community.}

\section{Introduction}

Since its discovery~\cite{fujisaka83}, chaotic synchronization is a
very active and important field of research~\cite{PikoBook}, with
applications in such diverse fields as secure
communications~\cite{first,roy_05}, semiconductor lasers~\cite{roy01},
chemical reactions~\cite{hudson}, living systems~\cite{postonov} and
electrically coupled neurons {\it in vitro} \cite{llinas}.

The phenomenology of the synchronization transition (ST) is
particularly rich for spatially extended chaotic systems, where a
parallel with critical phenomena can be drawn~\cite{stc}.  In the last
decade, an ongoing research activity has been devoted to relate chaotic
STs to non-equilibrium phase transitions
\cite{baroni,baroni99,bagnoli01,PA02,gino,GLPT03,ginelli_2,munoz,droz,cencini_torcini,gade,rechtman,noi}.
Nowadays, it is well established that the synchronization of two
replicas of a chaotic system coupled via spatio-temporal
noise~\cite{baroni}, or via local transverse coupling~\cite{PA02}, can
be characterized as a continuous non-equilibrium transition from an
active to an absorbing phase. In systems with nearest-neighbor (NN)
couplings, it has been found that the STs belong to two universality
classes, depending on the local dynamical features, namely: directed
percolation (DP)~\cite{Gras,haye} or multiplicative noise
(MN)~\cite{munoz_review}. These studies have been mainly performed for
coupled map lattices (CML), which are prototype models for systems
exhibiting spatio-temporal chaos~\cite{stc1}. An order parameter for
the STs is represented by the propagation velocity of information
$V_I$ \cite{TORCINI-POLITI-GRASSBERGER,cencini_torcini_pre}: in the
synchronized (desynchronized) state this quantity is zero (finite).
Moreover, for continuous maps, where the dynamics is dominated by
linear effects $V_I$ vanishes together with the {\it transverse
Lyapunov exponent}. In this case the critical indexes associated to
the ST are of the MN kind \cite{PA02}. On the other hand, for
discontinuous (or {\it quasi-discontinuous}~\cite{nota0}) maps, due to
the predominance of nonlinear effects, the replicas synchronize for
definitely negative transverse Lyapunov exponents
\cite{TORCINI-POLITI-GRASSBERGER,baroni}. For these maps
the ST is now characterized by critical properties typical of DP, 
analogously to what happens for cellular automata~\cite{G99,rechtman}.
These findings have been also confirmed by the analysis of
stochastic models mimicking the synchronization of continuous/discontinuous
maps ~\cite{GLPT03}.

Recently, synchronization has been also studied for chaotic systems
presenting long-range interactions, which are relevant to many real
contexts, such as disease spread via aviation traffic~\cite{geisel},
neuron populations~\cite{neurosynch}, Josephson junctions~\cite{jose}
and cardiac pacemaker cells~\cite{cuore}.

Typically, long-range interactions are introduced by considering CMLs
with coupling decaying as a
power-law~\cite{PV94,told,anteneodo,cencini_torcini}. In such a model,
when the local dynamics is dominated by nonlinear mechanisms, the STs
have been shown to belong to a family of universality classes known as
{\it Anomalous Directed Percolation} (ADP)~\cite{noi,haye_rev_lr}. ADP
has been previously identified for epidemic spreading whenever the
infective agent can perform unrestricted L{\'e}vy
flights~\cite{HH98,janssen}.  Such processes, originally introduced in
Ref.~\cite{mollison77}, can be modeled by assuming, e.g. in $d=1$,
that the disease propagates from an infected site to any other 
with a probability $P(r) \sim r^{-(1+\sigma)}$ algebraically decaying
with the spatial distance $r$, where $\sigma$ controls the interaction
range. Hinrichsen and Howard (HH) have numerically shown for a
stochastic lattice model (generalizing directed bond percolation) that
the critical exponents vary continuously with $\sigma$~\cite{HH98}.
These findings confirm previous theoretical results~\cite{janssen}
indicating that usual DP should be recovered for sufficiently
short-ranged coupling (namely, $\sigma > \sigma_c \equiv 2.0677(2)$)
and that a mean-field description should become exact for $\sigma <
\sigma_m \equiv 0.5$ (for a recent and exhaustive review see
Ref.~\cite{haye_rev_lr}).

In this paper, we first reconsider, by performing more accurate
estimations of the critical exponents, the results obtained for NN
interactions both for deterministic as well as for stochastic local
dynamics~\cite{gino,GLPT03}. Then we focus on the
synchronization transition in systems with power-law coupling. In this
context we extend previous analysis for discontinuous deterministic
maps~\cite{noi} to a stochastic version of the model, and we confirm
that whenever the STs are driven by nonlinear effects the associated
critical properties belong to the anomalous directed percolation
universality classes.  Finally, we present a first evaluation of the
critical properties of the synchronization transitions for continuous
chaotic maps with power-law interactions.

The paper is organized as follows: the employed deterministic and
stochastic models are introduced in the next section, while
Section~\ref{sec:3} is devoted to the methods used to study their
critical properties. In Section~\ref{sec:4}, STs are re-examined for
nearest-neighbor interacting systems; while Sec.~\ref{sec:5} is
focused on power-law coupled systems. Conclusions and perspectives
are discussed in Sec.~\ref{sec:concl}. The Appendix presents a derivation of
the field equation, known to reproduce ADP critical phenomena, for the
stochastic model with power-law interactions.

\section{Models}
\label{sec:2}

In this paper we investigate the synchronization between two 
replicas of CMLs transversally coupled according to the following scheme
\begin{eqnarray}
x_i(t+1) &=& (1-\gamma) F(\tilde{x_i}(t))+\gamma F(\tilde{y_i}(t))
\nonumber\\
y_i(t+1) &=& (1-\gamma) F(\tilde{y_i}(t))+\gamma F(\tilde{x_i}(t))\,,
\label{eq:cml}
\end{eqnarray}
where $i=1,\dots,L$ is the discrete spatial index with $L$ denoting the
system size; $x_i(t),y_i(t) \in [0:1]$ are the state variables and
$F(x)$ determines the local dynamics on each site of the lattice (see
below for its specification). The parameter $\gamma$ sets the strength
of the site-by-site coupling between the two replicas.

The variables $\tilde{z_i} \in \{\tilde{x_i}, \tilde{y_i}\}$ indicate
spatially averaged quantities. In particular, two different kind of spatial 
averages are introduced to reproduce short and long-ranged interactions. 
The former represents a discretized version of spatial diffusion among
nearest neighbour sites and reads:
\begin{equation}
\tilde{z}_i= (1-\epsilon) z_i +\frac{\epsilon}{2} (z_{i-1}+z_{i+1})\,, 
\label{eq:diffusive}
\end{equation}
where $\epsilon$ measures the intensity of the diffusive interaction
within each replica.  The latter is obtained
by considering a coupling  decaying as a power law
\cite{PV94,anteneodo,cencini_torcini}, i.e.
\begin{equation}
\tilde{z}_i=(1-\epsilon) z_i + \frac{\epsilon}{\eta(\sigma)}
\sum_{k=1}^{L^\prime}\frac{z_{i-k}+z_{i+k}}{k^{1+\sigma}}\,,
\label{eq:power_law}
\end{equation}
where $\sigma$ tunes the interaction range (the rationale for
defining the exponent as $1+\sigma$ will become clear in
Sec.~\ref{sec:5}). For $\sigma\to \infty$, Eq.~(\ref{eq:power_law})
reduces to the diffusive case (\ref{eq:diffusive}), while for
$\sigma=-1$ it corresponds to the mean-field coupling, usually
employed in the study of globally coupled maps~\cite{gcm}.
Actually, as discussed in Ref.~\cite{cencini_torcini}, the large scale
properties of the system should coincide with those at $\sigma=0$ for
any $\sigma \in [-1:0]$. Since the sum extends up to $L^\prime =
(L-1)/2$ the model is well defined only for odd $L$-values, and
$\eta(\sigma)=2\sum_{k=1}^{L^\prime} k^{-(1+\sigma)}$ is a
normalization factor to ensure that $\tilde{z}\in [0:1]$.  For both
kinds of coupling (\ref{eq:diffusive}) and (\ref{eq:power_law})
periodic boundary conditions are assumed and the diffusive coupling
$\epsilon$ is set to $2/3$.  Notice that the results do not seem to
depend on the chosen value of $\epsilon$.
\begin{figure*}[ht!]
\begin{center}
 \includegraphics[width=4.0cm]{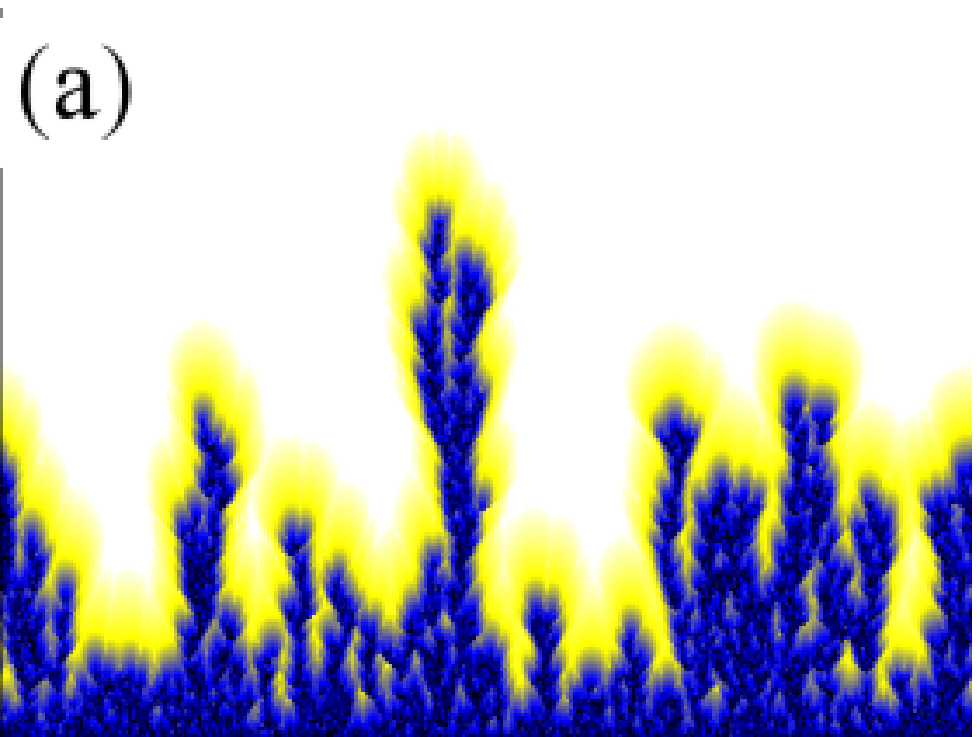}
 \includegraphics[width=4.0cm]{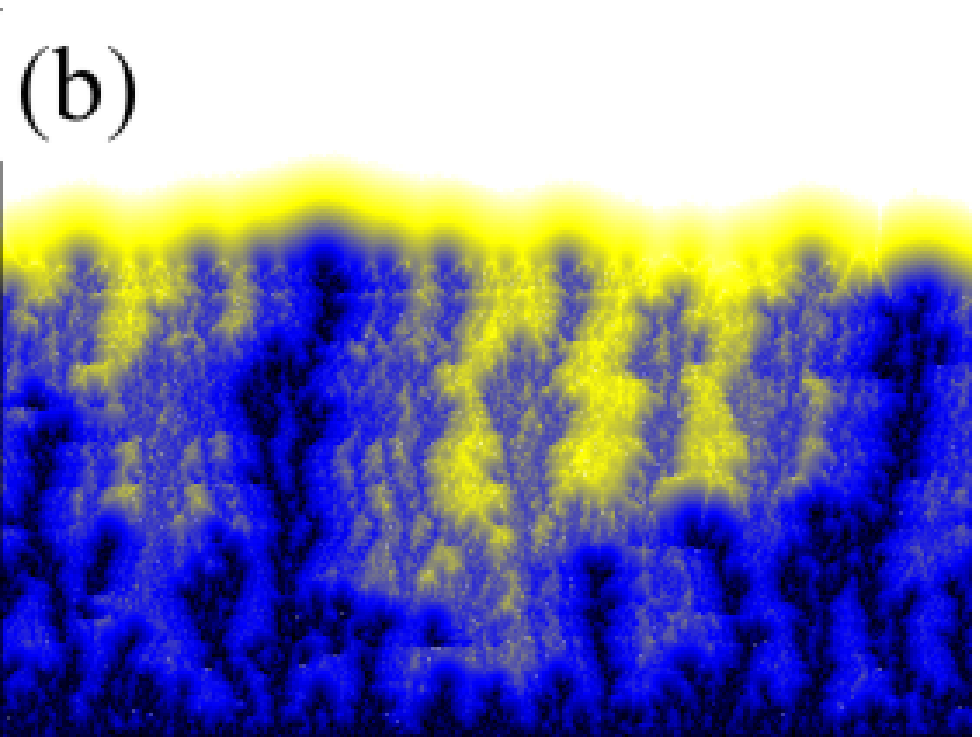}
 \includegraphics[width=4.0cm]{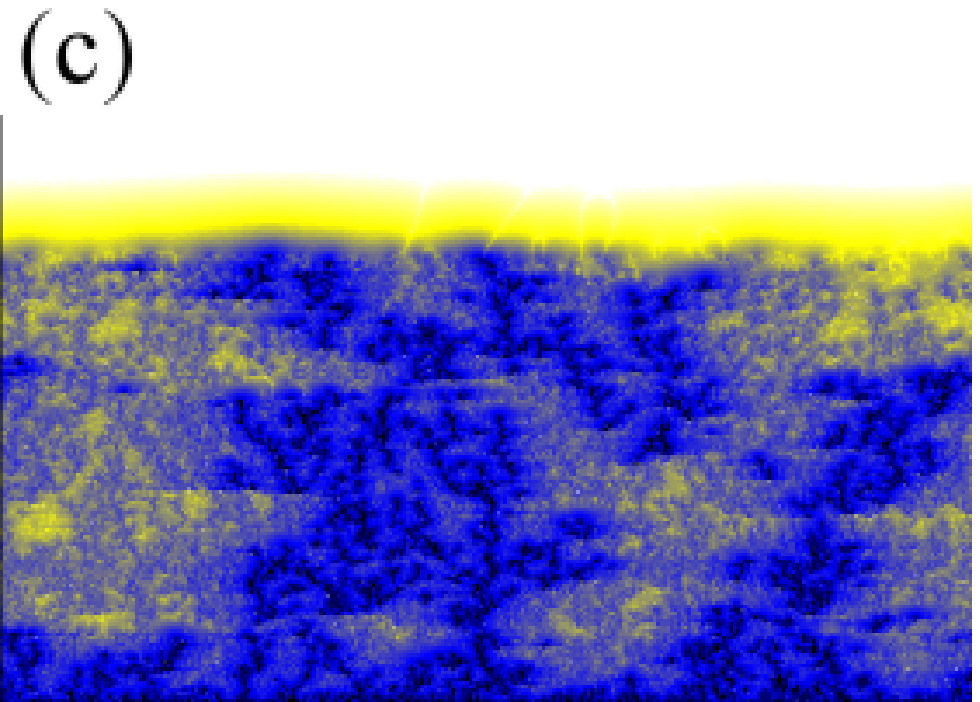}
 \includegraphics[width=4.0cm]{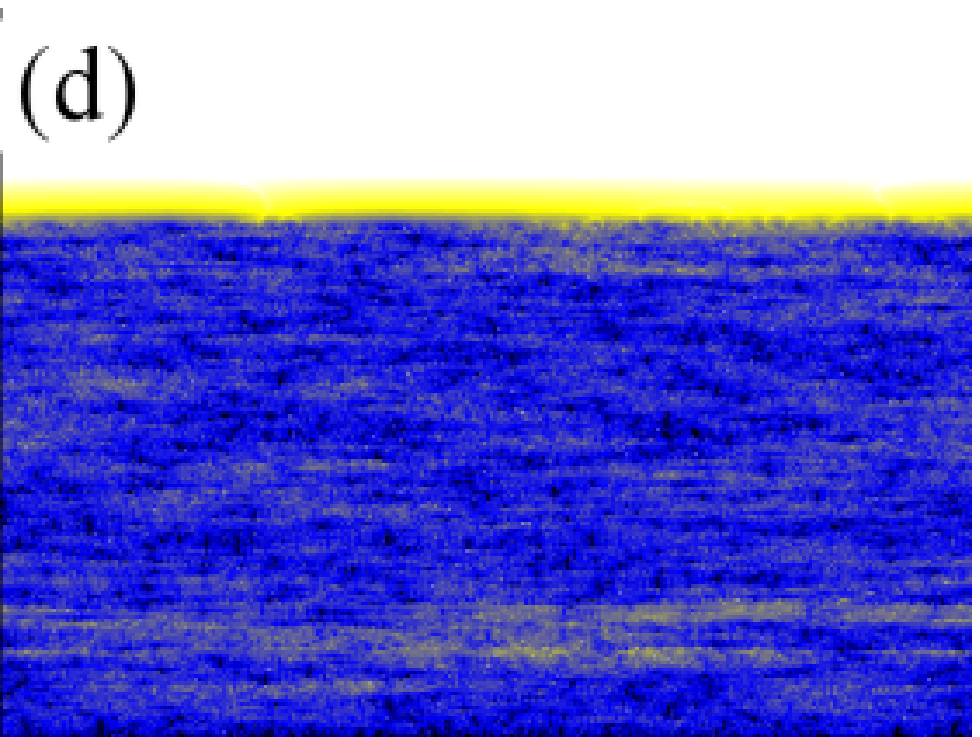}\\
 \includegraphics[width=4.0cm]{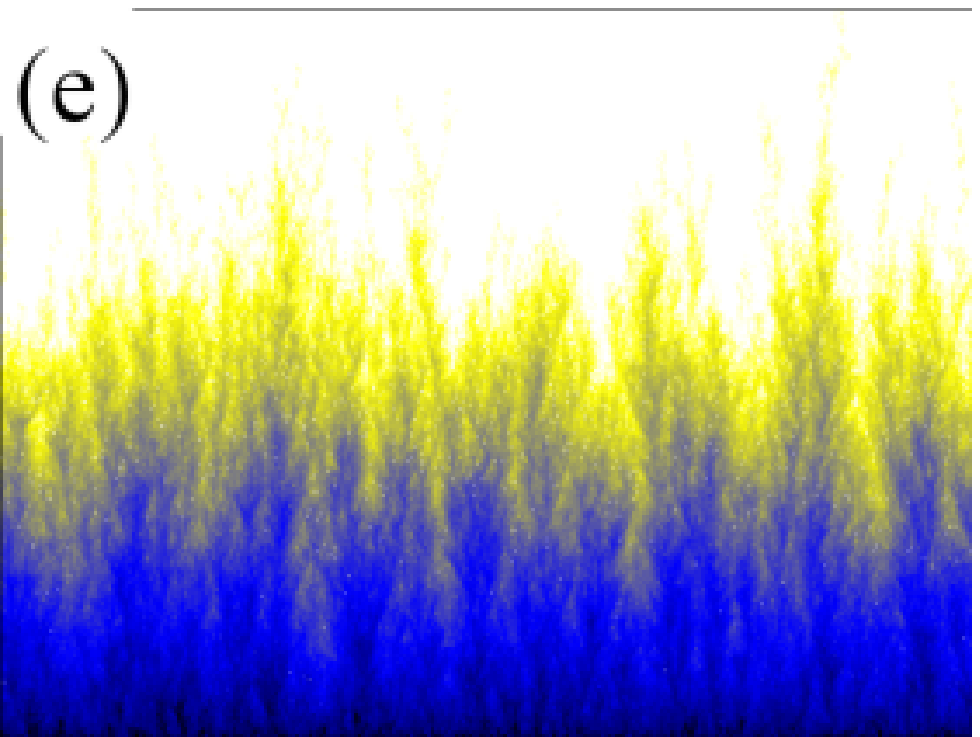}
 \includegraphics[width=4.0cm]{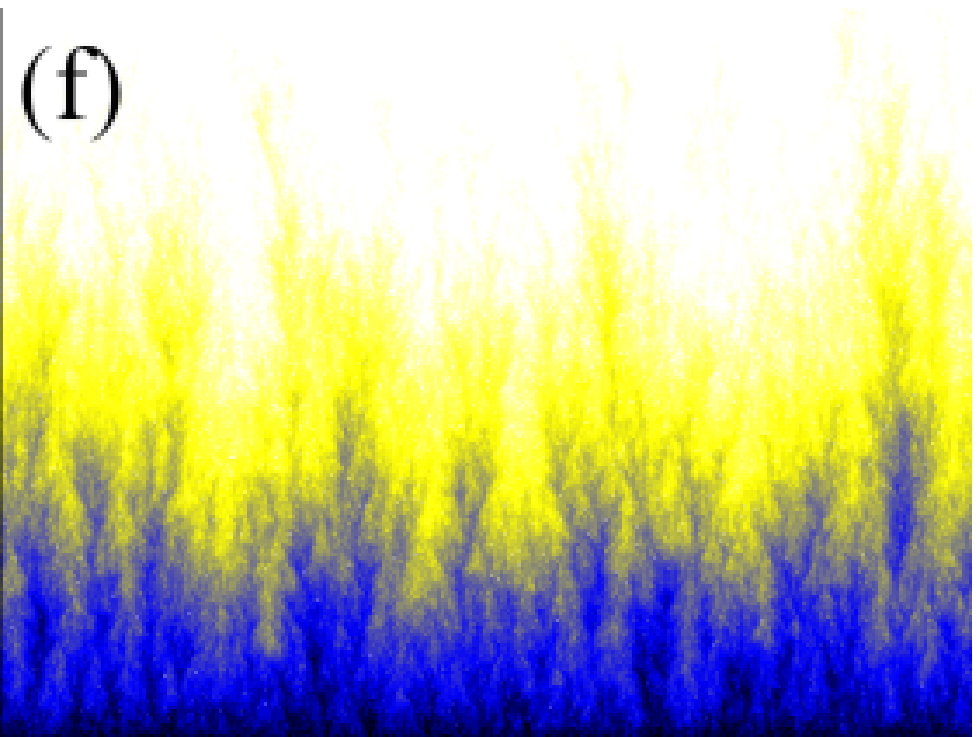}
 \includegraphics[width=4.0cm]{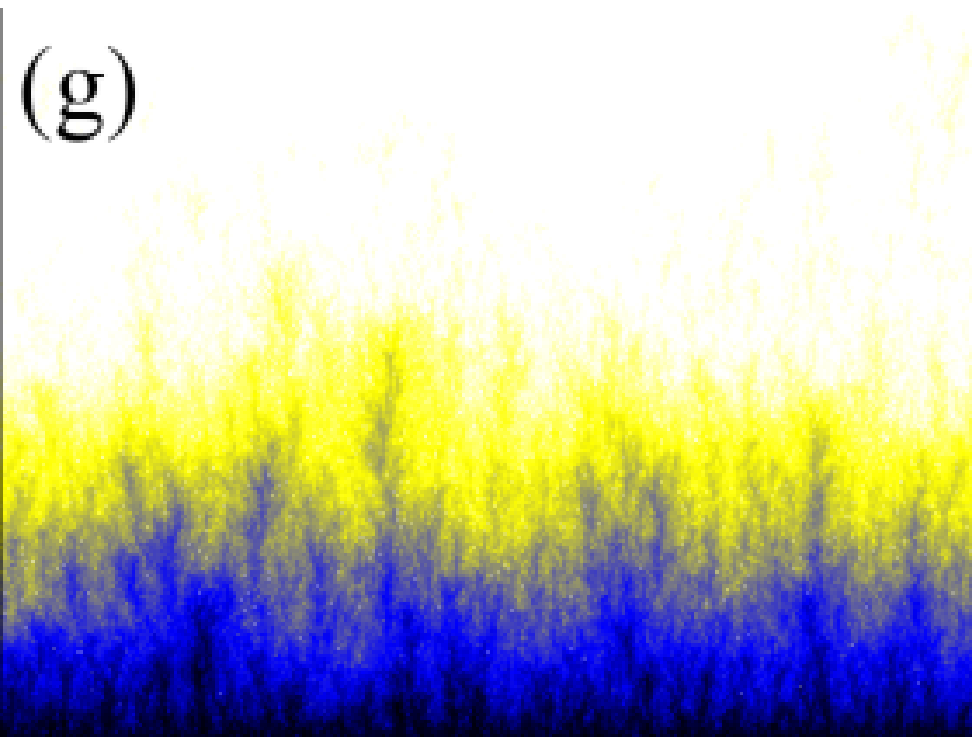}
 \includegraphics[width=4.0cm]{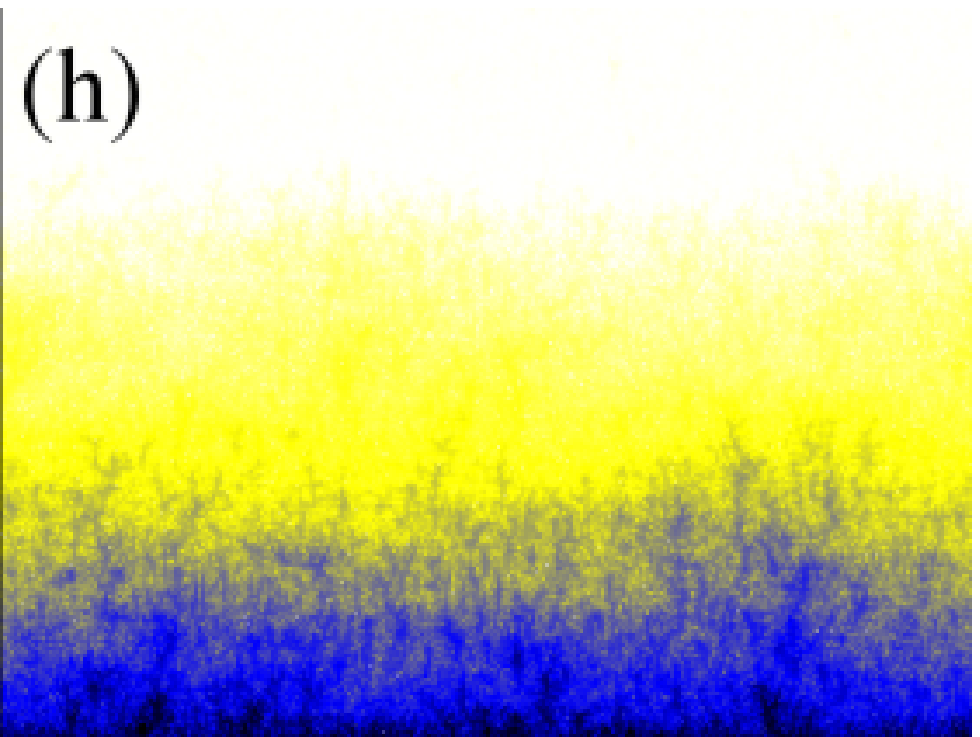}
\end{center}
\caption{\label{fig1} (Color online) The spatio-temporal evolution of
  the synchronization error for power-law coupled maps is depicted for local dynamics 
  given by the Bernoulli map (upper row) and the tent map (lower
  row). Each column corresponds to different values of the 
  exponent $\sigma$: namely, $\sigma=5$ (panels (a) and (e));
  $\sigma=2$ (panels (b) and (f)); $\sigma=1.2$ (panels (c) and (g))
  and $\sigma=0.5$ (panels (d) and (h)). The horizontal axis
  represents space, while the vertical one, time. 
  Note that the patterns observed for NN interaction closely resembles
  those reported in (a) and (e) for discontinuous and continuous maps,
  respectively.  The color code represents the synchronization error 
  (in logarithmic scale), ranging from white ($w_i(t) \to 0$) to 
  black ($w_i(t) =1$), larger values correspond to darker
  tones. The results reported in the panels have been obtained by employing
  the modified model (\ref{eq:smartcoupling}) for $q=4$ and $M=5$. }
\end{figure*}

Several studies
\cite{cencini_torcini_pre,baroni,PA02,GLPT03,cencini_torcini} have
shown that the system dynamics is strongly influenced by the
continuity properties of the local map $F(x)$
\cite{TORCINI-POLITI-GRASSBERGER}.  In particular, the spatio-temporal
propagation of disturbances \cite{cencini_torcini_pre} and the
properties of synchronization~\cite{PA02} are noticeably different if
the map is continuous or (quasi) discontinuous.  For this reason,
throughout this work, we consider two different deterministic maps:
the Bernoulli map $F(x) = 2 x\, (\hbox{mod} \, 1)$, which is
characterized by a discontinuity making dominant the nonlinear
effects, and the tent map $F(x) = 1 - 2 |x-1/2|$, whose dynamical
properties are well captured by linear analysis.

The synchronization transition is observed while increasing the
coupling $\gamma$: above a critical value $\gamma_c$ the two replicas
synchronize onto the same chaotic trajectory. In other words, for
$\gamma \ge \gamma_c$ the synchronization error
$w_i(t)=|x_i(t)-y_i(t)|$ vanishes for sufficiently long times.

In parallel with Eq.~(\ref{eq:cml}), we also consider a stochastic
model which mimics the evolution of the synchronization error ({\it
difference field}) $w_i(t)$ in proximity of the ST.  Stochastic
dynamics is expected to deplete the role played by spatio-temporal
correlations, unavoidable in deterministic systems, and thus
should allow more accurate estimations of the critical
exponents.  

In particular, we consider a stochastic model able to reproduce the
main features of ST both for continuous and discontinuous maps. We
consider the so-called Random Multiplier (RM) model, originally
proposed in Refs.~\cite{gino,GLPT03}.  The quantity $w_i(t)$ defined
in each site of the chain evolves according to the following
stochastic dynamics:
\begin{equation}
w_i(t+1)= \left\{
\begin{array}{ll}
1, & \mbox{w.p.} \quad p=a\tilde{w}_i(t) \\
a\tilde{w}_i(t), & \mbox{w.p.} \quad 1-p
\end{array}
\right. \;\; {\rm if} \; \tilde{w}_i(t) > \Delta
\label{RMmodel1}
\end{equation}
\begin{equation}
w_i(t+1)= \left\{
\begin{array}{ll}
\tilde{w}_i(t)/\Delta, & \mbox{w.p.} \quad p=a \Delta \\
a\tilde{w}_i(t), & \mbox{w.p.} \quad 1-p
\end{array}
\right. \;\; {\rm if} \; \tilde{w}_i(t) \le \Delta
\label{RMmodel2}
\end{equation}
where $w_i(t) \in [0:1]$ and $\tilde{w}_i$ indicates the spatial
averages that depending on the nature of the coupling, NN or
long-range, is given by (\ref{eq:diffusive}) or (\ref{eq:power_law}).
The parameter $\Delta$ controls the nonlinear effects, while $a$ the
linear ones. For the local maps considered in this article $a$ can be
identified with $2 (1-2 \gamma)$ \cite{GLPT03}.  Therefore, to vary
the coupling $\gamma$ in the deterministic model amounts to modify $a$
in the stochastic one. Notice that, due to the relationship
between the parameters $\gamma$ and $a$, in the RM model to achieve
synchronization $a$ should be reduced. 
For sufficiently small (resp. large) $\Delta$ the model (\ref{RMmodel1}-\ref{RMmodel2}) 
mimics the dynamics of (quasi)-discontinuous (resp. continuous) maps, where nonlinear
(resp. linear) effects are predominant. Following Ref.~\cite{GLPT03}, in
order to reproduce the behaviors of the Bernoulli or of
continuous-like maps we fixed $\Delta =0$ or $\Delta = 0.2$,
respectively.

\section{Characterization of the synchronization transitions}
\label{sec:3}

The phenomenology of the synchronization transitions at varying the
coupling range and the type of local dynamics can be visualized by looking 
at Fig.~\ref{fig1}. The figure displays the difference field $w_i(t)$ for
various values of $\sigma$ with transversal coupling just above the synchronization transition
($\gamma \gtrsim \gamma_c$) for Bernoulli as well as for tent maps. 
The spatio-temporal evolution of $w_i(t)$ is strongly dependent not only
on the continuous/discontinuous nature of the local dynamics, but also
on the interactions range within each replica.  For the discontinuous
maps (Fig.~\ref{fig1}a-d), percolating structures are clearly visible
in the short range limit (i.e. large $\sigma$).  However, as the
range of the interaction increases (i.e. $\sigma$ decreases) these
spatial structures tend to be smoothed out and, finally, for
$\sigma\lesssim 0.5$, they are no more detectable.  When the local
dynamics is governed by a continuous map, the qualitative results
change completely (Fig.\ref{fig1}e-h). In this case, desynchronization
(resurgence) phenomena are possible within the already synchronized
areas and this leads to a less defined distinction between
synchronized and desynchronized regions. However, a disappearance of the 
spatial structures is once more observable by increasing
the interaction range.

The patterns observed in Fig.~\ref{fig1} suggest a link between
synchronization transitions and non-equilibrium phase transitions from
an {\it active} to an {\it absorbing} ({\it quasi-absorbing}) {\it
phase}~\cite{haye,munoz_review}. In particular, Fig.~\ref{fig1}a
recalls the patterns observed for Directed Percolation (DP)~\cite{Gras,haye}, 
a contact process usually employed in the description of epidemics spreading. 
DP has been the subject of active research in
the last 20 years and, notwithstanding exact analytical results
are still lacking, detailed numerical studies have been able to determine its
critical properties up to a noticeable accuracy~\cite{jensen}.  Quite
recently, also experimental measurements of the critical DP indexes
have been reported for a transition to spatio-temporal intermittency
in a quasi one-dimensional system consisting of a ring of ferrofluidic
spikes~\cite{Tedeschi}, and for a transition between two topologically
different turbulent states in a quasi two-dimensional layer of nematic
liquid crystals~\cite{Chate}.  Fig.~\ref{fig1}e resembles instead
surface roughening, indeed Kurths and Pikovsky~\cite{pikov_kurths}
have proved that within a linear framework the evolution of the 
difference field of coupled CMLs is described by a Kardar-Parisi-Zhang
equation~\cite{kpz} in the presence of a hard wall. This latter
equation can be put in directed relationship with the Multiplicative
Noise (MN) Langevin equation~\cite{munoz_review}, which reproduce 
critical properties of non-equilibrium pinning-depinning as well as
wetting-dewetting transitions~\cite{Hin1,ginelli_2}.

As mentioned in the Introduction, the above analogies are not merely qualitative, 
indeed several studies~\cite{baroni,PA02,munoz,GLPT03} have quantitatively shown
that, depending on the prevalence of linear (resp. nonlinear) effects
in the local dynamics, ST in diffusively coupled systems belongs to
the MN (resp. DP) universality class. In particular, by changing the
value of the local multiplier (a control parameter), one can pass
continuously from one class to the other~\cite{GLPT03}, thus
suggesting that both these non-equilibrium transitions can be
described within a single field-theoretic framework. This is confirmed
by further results reported in Ref.~\cite{ginelli_2} for a microscopic
model of wetting transitions and in Ref.~\cite{munoz} for  KPZ
equation with an attractive wall.  Indeed, for such processes the time
evolution of the local density of active sites $n(x,t)$ admits a {\it
formally} quite similar Langevin description :
\begin{equation}
\partial_t n=D_N\nabla^2 n +\tau n-\Gamma n^p+g(n) \xi\,.
\label{eq:langevin}
\end{equation}
Such equation recovers the Reggeon-Field Theory (RFT)
describing DP for $p=2$ and $g(n)\! \propto\!  \sqrt{n}$,
while the minimal model for MN can be obtained from Eq.~(\ref{eq:langevin})
whenever $g(n)\propto n$. In the synchronization context, $n$ represents
the coarse grained synchronization error obtained by averaging
$w_i(t)=|x_i(t)-y_i(t)|$ over a suitable space-time cell. The
parameters entering Eq.~(\ref{eq:langevin}) are the diffusion
coefficient $D_N$ (corresponding to $\epsilon$); $\tau$ that measures
the distance from the critical point (i.e. from the synchronization
threshold $\gamma_c$) and the amplitude of the nonlinear term
$\Gamma$.  Finally, $\xi$ is a zero-average $\delta$-correlated (in
space and time) Gaussian noise field with unit variance.

In Ref.~\cite{janssen,HH98} the RFT has been extended to include long-range 
interactions, leading to the following stochastic equation
\begin{equation}
\partial_t n=D\nabla^2 n +D_A\nabla^\sigma n + \tau n-\Gamma n^2+g(n)
\xi\,,
\label{eq:langevin1}
\end{equation}
which generalizes (\ref{eq:langevin}) through the addition of an
anomalous diffusion term with coefficient $D_A$ and range of
interaction parametrized by $\sigma$.  For $g(n)\propto \sqrt{n}$,
Eq.~(\ref{eq:langevin1}) describes the ADP
universality class while, to the best of our knowledge, such an
equation has never been studied in the context of {\it anomalous} MN;
i.e for $g(n)\propto n$.

Recently in Ref.~\cite{noi} we have numerically shown that STs
observed for model (\ref{eq:cml}) with power law coupling
(\ref{eq:power_law}) and equipped with discontinuous maps (namely,
Bernoulli maps) belong to ADP universality class.  In this context the
extension of the analysis to the stochastic models
(\ref{RMmodel1},\ref{RMmodel2}) would be particularly interesting,
for two reasons. Firstly, the model
(\ref{RMmodel1},\ref{RMmodel2}) is known to reproduce quite well 
the phenomenology of synchronization for systems like 
(\ref{eq:cml}) with short-range coupling~\cite{GLPT03}. Second,
after a suitable coarse-graining, such models for $\Delta=0$ are effectively described
by the Langevin equation derived for RFT, providing further support
to the connection with DP~\cite{GLPT03}.  Here, in the Appendix, we show that this
kind of mapping can be extended also to the power-law interacting
model and, in this case, Eq.~(\ref{eq:langevin1}) is recovered.
Furthermore, the aim of the present work is to extend the studies
reported in Ref.~\cite{noi} also to continuous maps.

Before reporting a detailed analysis of the synchronization transition
for the above defined models, let us introduce the exponents employed
to characterize STs as non-equilibrium transitions. These are defined through 
the spatially averaged
synchronization error $\rho_\gamma(t)=\sum_i w_i(t)/L$ which, at
sufficiently long times, vanishes whenever a complete synchronization
is achieved, i.e., $\rho^\ast_\gamma= \lim_{t\to \infty}
\rho_\gamma(t)=0$ for $\gamma>\gamma_c$; whilst it remains finite at
any time in the desynchronized state, i.e. for $\gamma<\gamma_c$.

The order parameter $\rho_\gamma(t)$ allows us to define the critical
exponents $\theta$, $\beta$ and $z$ which characterize the transition:
$\theta$ is the exponent that rules the temporal scaling of the order
parameter at the critical point $\gamma = \gamma_c$, namely
\begin{equation}
\rho_{\gamma_c}(t) \sim t^{-\theta}\,.
\label{delta}
\end{equation}
For $\gamma < \gamma_c$, $\rho^\ast_\gamma\neq 0$ and one has
\begin{equation}
\rho^\ast_\gamma \sim (\gamma_c-\gamma)^\beta 
\label{beta}
\end{equation}
which defines the critical exponent $\beta$. Finally, the
dynamical exponent $z$ can be defined in terms of the finite-size
scaling relation, valid at the critical point:
\begin{equation}
\rho_{\gamma_c}(t) \sim L^{-\theta z} f(t/L^z)\,.
\label{finite_size}
\end{equation}
These three indexes are sufficient to fully characterize the
transition, since from their knowledge all the other critical exponents
can be derived. In particular, the exponent $\nu_\perp$ 
(resp. $\nu_\parallel$) ruling the divergence
of spatial (resp. temporal) correlation lenght at the critical
point is given by
\begin{equation}
\nu_\perp = \frac{\beta}{\theta z }     
\qquad \left({\rm resp.} \qquad \nu_\parallel = \frac{\beta}{\theta} \quad \right)
\label{corr_exp}
\end{equation}
A detailed description of the numerical estimation of $\theta$, $\beta$ and
$z$ is reported in the following sections.

\section{Short-range interactions}
\label{sec:4}

\begin{table}[b!]
\centering
\begin{tabular}{||c|c|c|c||} \hline 
      & $\theta$ & $\beta$ & $z$ \\\hline \hline
Bernoulli  &   0.159(1)  &   0.27(1) &  1.58(4) \\
RM ($\Delta=0$) & 0.1595(4)  &  0.276(2) &  1.58(2)  \\
DP \cite{jensen}  &   0.159464(6) &   0.276486(6)   & 1.580745(6)  \\
\hline
\hline
Tent  &    1.275(15)  &  1.70(8) & 1.5(1)\\
RM ($\Delta=0.2$) &  1.13(3)  &  1.67(3) & 1.53(6) \\
MN \cite{grinstein} & 1.10(5) &  1.70(5) & 1.53(7)   \\
MN \cite{kissinger} &  1.184(10) &  1.776(15) & ---   \\
\hline
\end{tabular}
\caption{Critical exponents for nearest-neighbor coupled
tent and Bernoulli maps as well as for the corresponding stochastic models,
Also the best estimations of the critical indexes for DP 
and MN universality classes are reported.}
\label{tabNN} 
\end{table}

\begin{figure*}[th!]
\includegraphics[width=1\textwidth,angle=0]{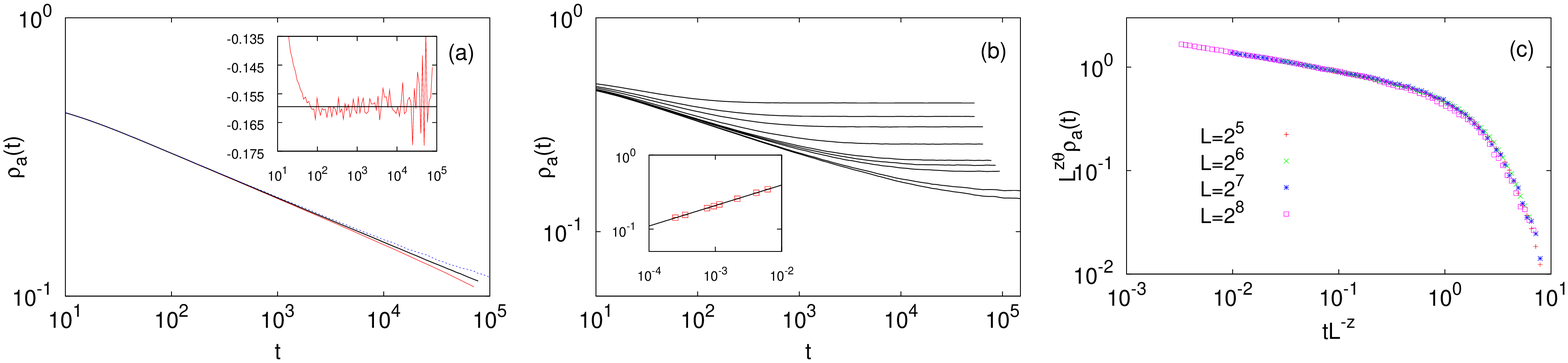}
\includegraphics[width=1\textwidth,angle=0]{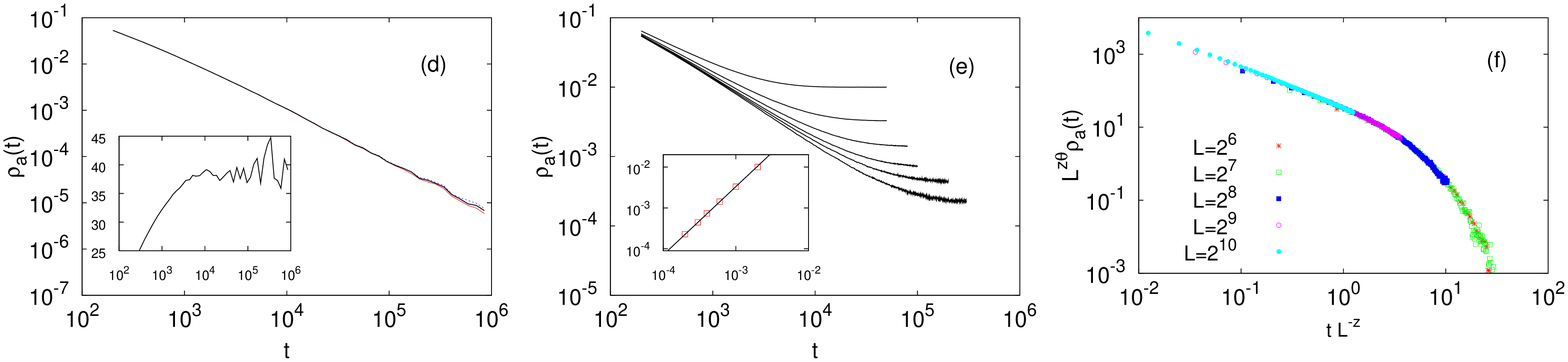}
\caption{\label{stoc_dp_nn} (top) Scaling relationships for the
  critical exponents for the RM model (\ref{RMmodel1}-\ref{RMmodel2})
  with NN coupling for $\Delta=0$: (a) $\rho_a(t)$ vs $t$ is displayed for
  $a=0.60620$, $0.60615$, $0.60610$ with $L=2^{19}$ maps and averaging
  over $5\times 10^2$ different realizations, the inset shows the logarithmic
  derivative $d\ln(\rho_a(t))/d\ln(t)$; (b) various curves 
 $\rho_a(t)$ vs $t$ are reported for $a < a_c$ revealing the saturation to $\rho^*_a$; 
  in the inset $\rho^*_a$ vs $(a-a_c)$ is reported together with the 
  best fit giving the $\beta$ estimation, here $L=2^{19}$ maps and the average 
  is over $500$ realizations; (c) data collapse based on (\ref{finite_size}),
  averaging over $10^{3}$-$10^4$ initial conditions. (bottom) Same as
  top for the RM model with $\Delta=0.2$: (d) analogous to (a) for
  $a=0.566938$, $0.566942$, $0.566945$ with $L=5\times10^5$ and
  averaging over $1-2\times 10^2$ realizations, here in the inset we report
  ${\rho_a}_c(t)\, t^\theta$ vs $t$; (e) same as (b) with
  $L=5\times10^5$ and averaging over $2\time10^2-1.2\times10^3$ initial conditions;
  (f) analogous to (c), averaging over $10^5$ initial
  conditions. The values of the estimated critical exponents are
  reported in Table~\ref{tabNN}.  }
\end{figure*}
 
Let us now consider the synchronization of two replicas of
diffusively coupled map lattices. 
We start with the case of coupled tent maps where the linear analysis
is sufficient to identify the critical point and we can thus limit to
study the behavior of the difference field $w_i(t)$ in the tangent
space. In particular, the norm of this field grows at a rate given by
the transverse Lyapunov exponent~\cite{PA02}:
\begin{equation} 
\lambda_{\bot} = \lim_{t\to \infty} || w_i(t) || =  \ln ( 1 - 2 \gamma ) +
\Lambda;
\label{TLE}
\end{equation} 
where $\Lambda$ is the maximal Lyapunov exponent of a single
CML. The synchronization transition occurs at the point where
$\lambda_{\bot}$ vanishes, locating the critical coupling to
\begin{equation} 
\gamma_c = \frac 1 2 \left( 1 - e^{ - \Lambda} \right).
\label{gc}
\end{equation} 

For this model, it has been suggested that $\gamma_c =
0.17615(5)$~\cite{PA02}, by extrapolating from finite size
measurements the asymptotic value for $\Lambda$.  We have estimated
the synchronization value from the scaling of the density $\rho_\gamma (t)$
for chains of length $L=2^{25}$ finding $\gamma_c = 0.17616(2)$, in
agreement with the previous estimation. The corresponding critical
exponents are reported in Table \ref{tabNN}, we observe that these
values are in reasonable agreement with the one reported in
literature for the MN universality class~\cite{grinstein,munoz_review}, 
apart for $\theta$, which is larger than the corresponding MN value.
This difference is probably due to the
presence of correlations in the dynamical evolution of these maps,
that limits the time range over which a critical scaling of the density 
$\rho_\gamma(t)$ can be found. In particular, in our analysis as well as in
Ref.~\cite{PA02}, the time interval where it is possible to observe a
clear power-law scaling is limited to $[10^2:4.5\times 10^4]$, since
at longer times the density (above and below $\gamma_c$) always
saturates to a (small) constant value.  As we show in the following,
much more accurate estimations can be obtained by replacing the tent
map with its stochastic version, thus reinforcing the hypothesis that
scaling laws are hindered by long-time dynamical correlations.

For coupled Bernoulli maps linear analysis fails in locating the
synchronization transition, therefore we are forced to directly
investigate the scaling of the density $\rho_\gamma(t)$.  In particular, for
$L=2^{17}$ we have estimated a critical value $\gamma_c=0.28752(1)$,
in agreement with Ref.~\cite{PA02} (see Table \ref{tabNN} for the
values of the critical exponents). In this case finite size effects 
and time correlations are less relevant, since good scaling can
be already observed in the time interval $[10^2:10^5]$ for much
shorter chains.

Let us now turn our attention to the Random Multiplier model
(\ref{RMmodel1},\ref{RMmodel2}), which reproduces the Bernoulli map
behavior for $\Delta=0$ and that of a generic continuous map for
$\Delta=0.2$ \cite{gino,GLPT03}.

For the RM model with $\Delta = 0$ the scaling laws are reported in
Fig. \ref{stoc_dp_nn}a-c. In this case the critical point is
at $a_c=0.60615(5)$ and the measured indexes are in perfect
agreement with the best estimates of DP's exponents. 
$\theta$ and $\beta$ coincide with the DP values up to the fourth and
third digit, respectively. The evaluation of $z$ is less accurate, since
it relies on a data collapse of finite size estimates (\ref{finite_size}).

In the case $\Delta = 0.2$, by considering the scaling in time of the
density (\ref{delta}) for $L=5\times10^5$ maps over a time span $t \in
[10^3 : 10^6]$, we have found that the critical point is located at
$a_c=0.566942(4)$. In particular, $\theta$ has been estimated by a
best fit to a power-law over more than two decades
(Fig.~\ref{stoc_dp_nn}d). The quality of the estimation of the other
two exponents can be appreciated from Fig.~\ref{stoc_dp_nn}e and
Fig.~\ref{stoc_dp_nn}f. The values of the critical indexes are in fairly good agreement
with those reported in Ref.~\cite{grinstein}, obtained by considering
a time and space discretized version of the corresponding Langevin
equation, i.e. Eq.~(\ref{eq:langevin}) with $g(n) \propto \sqrt{n}$.
It should be remarked that the exponents recently reported
in Ref.~\cite{kissinger}, for a suitable lattice model reproducing
KPZ-type interface growth, are slightly larger than ours, while the
$\theta$-value (namely, $\theta = 7/6 = 1.166{\bar 6}$) conjectured by
Droz and Lipowski~\cite{droz} is, within the error bars, consistent
with our estimation.

\begin{figure*}[th!]
\includegraphics[width=1\textwidth]{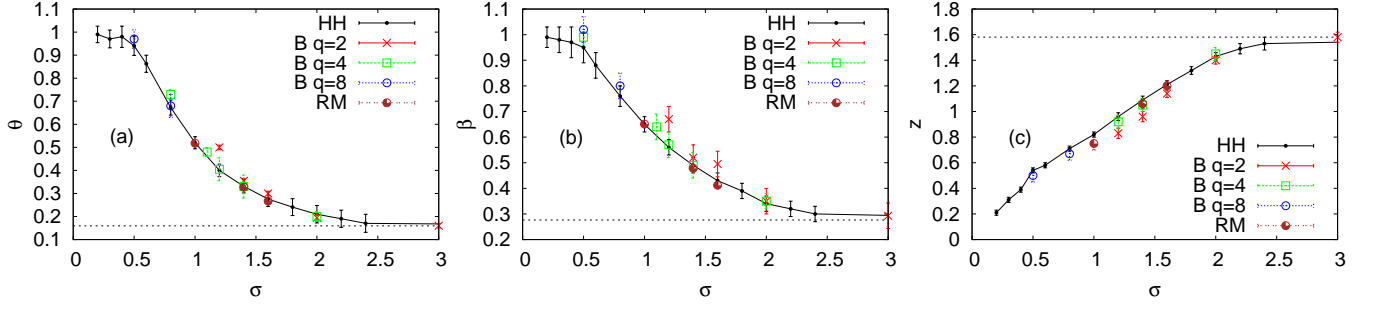}
\caption{(Color online) Critical exponents (a) $\theta$, (b) $\beta$ and (c) $z$ for
coupled Bernoulli maps and RM models (for $\Delta=0$) as a function of 
the long-range power exponent $\sigma$ compared with those obtained 
by HH in Ref.~\cite{HH98} for a stochastic lattice model.
Symbols are explained in the legend. The dotted horizontal line
represents the best estimate of the critical exponents for
usual DP, i.e. for nearest-neighbor spreading.}
\label{fig:adp}
\end{figure*}

\begin{table*}[t!]
\centering
\begin{tabular}{|c|ccc|ccc|ccc|ccc|ccc|}
\hline
 $\bm \alpha$ &  \multicolumn{3}{c|}{{\bf HH} } & \multicolumn{3}{c|}{{\bf B} $q=2$}
& \multicolumn{3}{c|}{{\bf B} $q=4$}& \multicolumn{3}{c|}{{\bf B} $q=8$}& \multicolumn{3}{c|}{{\bf RM} $q=4$}\\ 
\hline
   &$\theta$&$\beta$&$z$&$\theta$&$\beta$&$z$&$\theta$&$\beta$&$z$&$\theta$&$\beta$&$z$&$\theta$&$\beta$&$z$\\ 
 \hline
0.2& 0.99(4) & 0.99(4)&0.21(2)& \multicolumn{3}{c|}{-}&\multicolumn{3}{c|}{-} &\multicolumn{3}{c|}{-}&\multicolumn{3}{c|}{-}\\
0.5& 0.94(4) & 0.95(6)&0.54(2)& \multicolumn{3}{c|}{-}&\multicolumn{3}{c|}{-} &0.97(4)&1.02(5)&0.50(5)&\multicolumn{3}{c|}{-}\\
0.6& 0.86(4) & 0.88(5)&0.58(2)& \multicolumn{3}{c|}{-}&\multicolumn{3}{c|}{-} &\multicolumn{3}{c|}{-}&\multicolumn{3}{c|}{-}\\
0.8& 0.67(3) & 0.76(4)&0.71(2)&\multicolumn{3}{c|}{-} &0.73(2)& 0.91(5)&-     &0.68(5)&0.80(5)&0.67(5)&\multicolumn{3}{c|}{-}\\
1.0& 0.52(3) & 0.65(3)&0.82(2)& \multicolumn{3}{c|}{-}&\multicolumn{3}{c|}{-} &\multicolumn{3}{c|}{-}&0.52(1)&0.65(1)&0.75(5)\\
1.1&   -     &   -    &  -    & \multicolumn{3}{c|}{-}&0.48(2)&0.64(5)&-      &\multicolumn{3}{c|}{-}&\multicolumn{3}{c|}{-}\\
1.2& 0.40(3) & 0.56(3)&0.96(3)&0.50(1)&0.67(5)&0.83(4)&0.40(5)&0.57(4)&0.92(5)&\multicolumn{3}{c|}{-}&\multicolumn{3}{c|}{-}\\
1.4& 0.33(3) & 0.49(3)&1.09(3)&0.35(1)&0.52(5)&0.96(4)&0.33(5)&0.49(5)&1.05(5)&\multicolumn{3}{c|}{-}&0.33(1)&0.48(2)&1.06(4)\\
1.5&   -     &   -    &  -    & \multicolumn{3}{c|}{-}&\multicolumn{3}{c|}{-} &\multicolumn{3}{c|}{-}&\multicolumn{3}{c|}{-}\\
1.6& 0.27(3) & 0.43(3)&1.21(3)&0.30(1)&0.49(5)&1.14(4)&\multicolumn{3}{c|}{-} &\multicolumn{3}{c|}{-}&0.27(1)&0.41(1)&1.20(4)\\
1.8& 0.24(4) & 0.39(3)&1.32(3)& \multicolumn{3}{c|}{-}&\multicolumn{3}{c|}{-} &\multicolumn{3}{c|}{-}&\multicolumn{3}{c|}{-}\\
2.0& 0.21(4) & 0.34(3)&1.43(3)&0.19(2)&0.35(5)&1.40(3)&0.20(2)&0.35(3)&1.45(5)&\multicolumn{3}{c|}{-}&\multicolumn{3}{c|}{-}\\
2.2& 0.19(4) & 0.32(3)&1.49(4)& \multicolumn{3}{c|}{-}&\multicolumn{3}{c|}{-} &\multicolumn{3}{c|}{-}&\multicolumn{3}{c|}{-}\\
2.4& 0.17(4) & 0.30(3)&1.53(5)& \multicolumn{3}{c|}{-}&\multicolumn{3}{c|}{-} &\multicolumn{3}{c|}{-}&\multicolumn{3}{c|}{-}\\
2.5&  -      &   -    &  -    & \multicolumn{3}{c|}{-}&\multicolumn{3}{c|}{-} &\multicolumn{3}{c|}{-}&\multicolumn{3}{c|}{-}\\
3.0&  -      &   -    &  -    &0.16(1)&0.29(5)&1.58(4)&\multicolumn{3}{c|}{-} &\multicolumn{3}{c|}{-}&\multicolumn{3}{c|}{-}\\
5.0& 0.159538& 0.276  &1.58   & \multicolumn{3}{c|}{-}&\multicolumn{3}{c|}{-} &\multicolumn{3}{c|}{-}&\multicolumn{3}{c|}{-}\\
 \hline 
 \end{tabular}
\caption{Summary of the results for the critical exponents (also shown in Fig.~\ref{fig:adp}) 
corresponding to coupled Bernoulli maps and to chains of RM-models (with $\Delta =0$)
at various values of $\sigma$. Results taken from Ref.~\cite{HH98} by HH 
are also reported for comparison.}
\label{tab:adp}
\end{table*}

\section{Long-range interactions}
\label{sec:5}

In the presence of long-range interactions, accurate numerical
analysis of the synchronization transition requires huge time
costs, this can be appreciated by noticing that at each time step
${\mathcal O}(L^2)$ operations are involved (see the coupling
definition in (\ref{eq:power_law})). Simulation times become prohibitive as the
range of the interaction increases (i.e. the exponent $\sigma$
decreases), because finite size effects becomes more relevant and
larger sizes are needed. It is therefore fundamental to reduce the
CPU cost for each time step. As in Refs.~\cite{told, noi}, to
achieve such aim we consider the following coupling scheme
(\ref{eq:power_law}):
\begin{eqnarray}
\tilde{z_i}=(1-\epsilon) z_i + \frac{\epsilon}{\eta(\sigma)} 
\Big[(z_{i\!-\!1}+z_{i+1}) + \nonumber\\
\sum_{m=1}^{M} \frac{z_{i-j_m(q)}+z_{i+j_m(q)}}{(j_m(q))^\sigma} 
\Big] \, . \label{eq:smartcoupling}
\end{eqnarray}
where $j_m(q)=q^{m}\! -\!1$ and $M=\log_q (L/2)$ \cite{nota1}; the
model with full coupling (\ref{eq:power_law}) is recovered for
$j_m(q)=m$, $M=L\!-\!1$ and by substituting $\sigma \to 1+\sigma$ (see
below for more details). Clearly, the new choice is very convenient
because each updating step only requires $\mathcal{O}(L\log_q L)$
operations instead of $\mathcal{O}(L^2)$ needed for the fully coupled
case.  The parameter $q$ is typically chosen as $q=2,4$ and $8$, while
the normalization factor is given by $\eta(\sigma)=2 [1+ \sum_{m=1,M}
  (j_m(q))^{-\sigma}]^{-1}$.

The critical properties of the reduced model ($j_m(q)\!=\!q^m \!-\!1$)
with exponent $\sigma$ map into those of the fully coupled one
($j_m\!=\!m$) with exponent $\sigma_{fc}\!=\!\sigma\!+\!1$. This can 
be easily understood by noticing that the two versions of the model
should display the same critical behavior once the spatial interactions
scale analogously. For the modified model the coupling weight over
the interval $]j_m(q):j_{m+1}(q)]$, containing a single coupled site,
is simply given by
$$
\frac{1}{j_{m+1}(q)^\sigma} \sim \frac{1}{q^{(m+1)\sigma}} 
\,,$$ 
while over the same interval the weight for the fully coupled model 
can be estimated by evaulating the sum
$$
\sum_{k=j_m(q)+1}^{j_{m+1}(q)} \frac{1}{k^{\sigma_{fc}}} \sim
\frac{1}{q^{(m+1)(\sigma_{fc}\!-\!1)}} \,. . 
$$ 
By comparing the two
expressions it is thus clear that the two weights scale in the same
manner when $\sigma_{fc}\!=\!\sigma\! +\!1$, as we have also numerically
verified in Ref.~\cite{noi}.

Before discussing the numerical results we observe that the coupling
scheme (\ref{eq:smartcoupling}) can be easily implemented both for
coupled maps as well as for the stochastic model.

\subsection{Discontinuous maps and ADP universality class}

Synchronization transition of long-ranged coupled systems with
discontinuous maps has been previously studied in
Ref.~\cite{cencini_torcini}, as far as the self-synchronization of a
single chain is concerned, and in Ref.~\cite{noi}, where the
synchronization of two (transversely coupled) replicas is
discussed. Here we summarize the results obtained in Ref.~\cite{noi}
for Bernoulli maps coupled with reduced coupling scheme~\cite{nota2}
and extend the analysis to the stochastic RM model with $\Delta=0$ in
the same coupling conditions.

In order to observe clean scaling laws, we have employed the reduced
scheme with increasingly larger $q$-values for smaller $\sigma$.  This amounts
to chain lengths varying from $L=2^{16}$ for $\sigma=5.0$ up to
$L=2^{22}$ for $\sigma=0.5$. Analogously to the NN case,
due to the failure of linear analysis, the critical point $\gamma_c$
has been located by considering the scaling relation (\ref{delta}),
in particular from the critical power-law decay the exponent $\theta$
can be estimated. The
critical index $\beta$ can be derived by considering the asymptotic
values $\rho_\gamma^*$ below the transition ($\gamma < \gamma_c$) as in
(\ref{beta}), finally $z$ is estimated by data collapse based on the
finite-size relation \eqref{finite_size} at $\gamma = \gamma_c$.  
As shown in Ref.~\cite{noi}, the
quality of the scaling behavior is very good.

Table~\ref{tab:adp} and Fig.~\ref{fig:adp} summarize our findings and
present the new critical exponents estimated for the RM model
(\ref{RMmodel1})-(\ref{RMmodel2}), where we have employed the coupling
scheme (\ref{eq:smartcoupling}) with $q=4$ and $M=7$ (for $\sigma=1.4$
and $1.6$) and $M=9$ (for $\sigma=1.0$).  As one can appreciate from
Table~\ref{tab:adp} the estimated errors are smaller for the RM than
for the Bernoulli map.
In table~\ref{tab:adp} we also report the critical indexes found by
Hinrichsen and Howard~\cite{HH98} for the Anomalous DP universality
classes. As one can see the results obtained for the Bernoulli map
approach those of HH for increasing $q$, i.e. for chains with larger
sizes.  This tendency suggests that the observed discrepancies are
mainly due to finite size effects, which however seem to be less
severe for the stochastic model. Indeed by considering the RM model 
with chain sizes $L \sim 3 \times 10^4 - 5 \times 10^5$ we found that 
the exponents coincides, 
within the error bars, with those reported by HH. 

According to the analysis reported in Ref.~\cite{janssen,HH98}, 
the upper critical dimension $d_c$ is $2 \sigma$ and therefore the
mean-field regime in one dimensional chain should establish below
$\sigma_m\equiv 0.5$. The corresponding critical exponents are 
$\theta_{MF}=\beta_{MF}=1$ and $z_{MF}=\sigma$ (in accordance the 
temporal and spatial correlation length exponents are given by 
${\nu_\parallel}_{MF}=1$ and ${\nu_\perp}_{MF}=1/\sigma$).
Therefore, the long-range nature of the interactions
is reflected in the mean-field regime only by the scaling
of the spatial correlations, since for short-range interactions
the exponents coincide apart $z_{MF}=2$ (and correspondingly
${\nu_\perp}_{MF}=1/2$). Our data reported at
$\sigma=0.5$ for the Bernoulli maps confirm the mean-field
expectations, but due to computational
limitations we could not explore smaller $\sigma$-values.

\begin{table}[b!]
\begin{center}
\begin{tabular}{|c|c|c|c|c|c|c|c|} 
\hline $\sigma\;[q]$ & 3.0 [2] & 2.0 [4]& 1.4 [4]& 1.2 [4] & 0.8 [8] &
0.5 [8] \\ \hline $\delta$ & 0.93(6) & -0.13(9) & -0.04(12)& -0.03(10)
& -0.04(9)&0.03(9)\\ \hline
\end{tabular}
\end{center}
\vspace{-0.5truecm}
\caption{Scaling relation (\ref{hyper}) for various values of
$\sigma$, for each measurement the corresponding basis $q$ is
reported.  For the RM-model with $q=4$ we obtained: for $\sigma=1.0$
$\delta=-0.03(2)$, for $\sigma=1.4$ $\delta=-0.04(4)$ and for $\sigma=1.6$
$\delta=-0.05(4)$. The measured $\delta$ are compatible with zero in the
range $[0.5:2.0677]$. See text for details.}
\label{tab1}
\end{table}

For ADP, it has also been shown~\cite{janssen,HH98} 
that the following hyper-scaling relation holds
\begin{equation}
\delta = 1 -\sigma +(1 - 2 \theta) z \equiv 0\,;
\label{hyper}
\end{equation}
in the $\sigma$-range $[\sigma_m:\sigma_c]$. 
The $\sigma_c$ value at which the behaviour of the
system should cross over to usual DP can be directly estimated 
by inserting the estimated values of the DP esponents in
Eq.(\ref{hyper}).  Quite astonishingly the cross over
takes place at $\sigma_c \equiv 2.0677(2) > 2$, as
suggested also by field theoretic arguments ~\cite{janssen,HH98}.
As shown in Table~\ref{tab1} the relation (\ref{hyper}) is fulfilled
within the error bars for the Bernoulli coupled maps as well as for
its stochastic version. Notice that, $\delta$ departs from zero only
at $\sigma=3.0 > \sigma_c$, where usual DP scalings are expected. 
Finally, in the Appendix we show that, after a suitable spatio-temporal
coarse-graining, the stochastic model can be effectively described by
a Langevin equation of the form (\ref{eq:langevin1}) corresponding
to the field theoretic description associated to ADP. 
These results further reinforce the parallel between ST induced by nonlinear
effects, in the presence of power-law interactions, and ADP.

\subsection{Continuous maps}
\begin{figure}[b!]
\includegraphics[width=.49\textwidth]{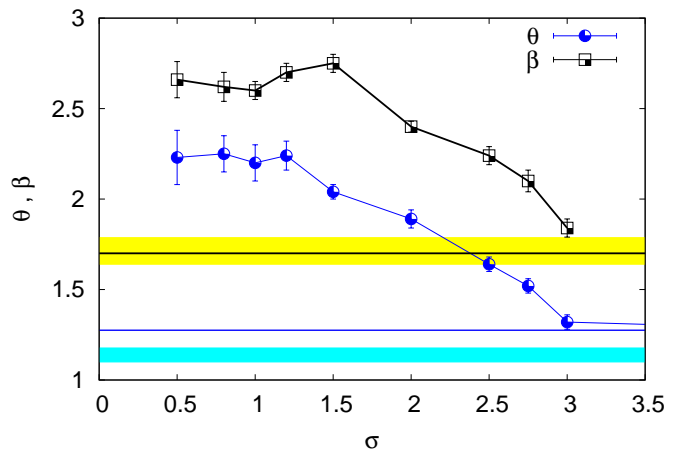}
\caption{\label{tent-delta} $\theta$ and $\beta$ as a function of
  $\sigma$ for the tent map with power-law coupling. The shaded upper
  (resp. lower) region corresponds to the best estimate of $\theta$
  (resp. $\beta$) reported in literature
  \cite{munoz_review,kissinger,droz} for the MN class in the case of
  short range interactions.  The thick upper (resp. lower) horizontal
  lines to the corresponding estimates for coupled tent maps with NN
  coupling reported in Table \ref{tabNN}.  The employed system sizes
  vary between $L=2^{19}$ and $L=2^{21}$ with $q=4$ for the long-range
  models. For systems with NN interactions we used sizes $L=2^{25}$
  for the $\theta$-estimation and $L=2^{18}$ for $\beta$.}
\end{figure}

\begin{figure}[t!]
\includegraphics[width=.49\textwidth]{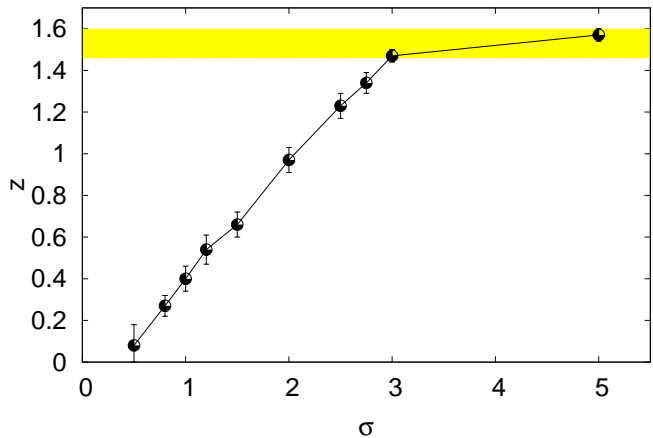}
\caption{\label{tent-zeta} $z$ vs $\sigma$ for the tent map with
power-law coupling. The shaded region corresponds the best estimate
of $z$ reported in literature \cite{munoz_review}
for the MN class in the case of short range
interactions. For the finite-size scaling chains of length
$L=2^4 - 2^{15}$ have been employed. }
\end{figure}

Let us finally consider two coupled replicas with local dynamics given by the tent map. 
In this case the critical point can be estimated by the vanishing of the transverse Lyapunov 
exponent accordingly to the expression (\ref{gc}). Furthermore, we have independently 
evaluated $\gamma_c$ by examining the critical behaviour of the density $\rho_\gamma(t)$. 
Usually the Lyapunov approach was able to locate $\gamma_c$ within a precision
of $1-2 \times 10^{-4}$ or even better.

In the present case, to simplify the analysis we keep the parameter
$q=4$ and we analyze system sizes ranging from $L=2^{19}$ to $L=2^{23}$.
The results for the exponents $\theta$ and $\beta$ are
shown in Fig.~\ref{tent-delta} while $z$ in Fig.~\ref{tent-zeta}, 
we observe that these critical
indexes for $\sigma \ge 3$ tend to the 
results reported for the MN universality class (see Table \ref{tabNN}). 
For smaller $\sigma$-values $\theta$ and $\beta$ (resp. $z$)
increase (resp. decreases) and appear to saturate (resp. to vanish) 
for $\sigma \le 1$. 
While the numerical values are completely different from the
ADP ones, the general trends are analogous
to those observed for coupled Bernoulli maps. This suggests
that in the present situation the mean-field value for the
indexes are $\theta_{MF} \sim 2.2(1)$, $\beta_{MF} \sim 2.6(1)$
and by assuming a linear dependence among $z$ and the power-law
exponent one finds that
$z_{MF} \sim 0.65(5) \times (\sigma-\sigma_0)$, with $\sigma_0 \sim 0.40(5)$.
At variance with the mean-field behaviour of the ADP case,
$\theta_{MF}$ and $\beta_{MF}$ do not coincide, thus 
suggesting that ${\nu_\parallel}_{MF} \sim 1.18(10)$
and $z$ seems to vanish already for $\sigma \sim 0.40(5)$.
Finally, from the numerical estimates of $z$ we
can conjecture that ${\nu_\perp}_{MF} \sim 1.8(5) / (\sigma-\sigma_0)$.
Apart from the poor quality of the estimates, it is important to stress
that, analogously to ADP, the presence of long-range interactions 
reflects mainly in the scaling of the spatial correlations. 
The mean-field limit for systems with multiplicative 
noise has been examined recently only for short-range couplings~\cite{gin04,cola05}. 
The authors of the papers agree on the values of the correlation exponents 
${\nu_\parallel}_{MF}=1$ and ${\nu_\perp}_{MF}=1/2$,  which implies 
$z_{MF}=2$ and $\theta_{MF}=\beta_{MF}$. 
However, the value of $\theta_{MF}$ seems to depend on the approximation employed
to derive the mean-field. In particular, in \cite{gin04,priv} a value
$\theta_{MF}=5/3$ is reported, while in \cite{munoz_review,cola05} is
predicted that  $\theta_{MF}$ is a nonuniversal scaling exponent dependent on noise 
amplitude, diffusion coefficient, and non-linear term. 
Therefore the mean-field analysis is still unclear for the multiplicative
noise case with usual diffusion (as discussed also in \cite{cola05}),
moreover a study for the fractional diffusion case is still to be addressed.

The analysis of the corresponding stochastic model, i.e.
the RM model with $\Delta =0.2$ with power-law coupling, reveal
enormous fluctuations in the asymptotic behavior of the
density induced by abrupt synchronizations of the whole chain. 
Thus making impossible any estimation of critical
scaling laws, even by employing quite long chains (namely, $L=2^{23}$). 
Our analysis cannot rule out the possibility that for the RM model
the transitions became discontinuous.  Similarly to what done in 
Ref. \cite{gin05}, a detailed analysis of the scaling of the synchronized clusters 
with the system size should be performed to address this 
point, but this goes beyond our scopes.
Moreover, it is quite astonishing that the analysis of the deterministic 
case was somehow clearer. We can conjecture that the correlations induced 
by the deterministic evolution of coupled tent maps prevent 
the abrupt synchronization of large islands within the
chain. However, this is just a working hypothesis to
be investigated in the future.

\section{Conclusions}
\label{sec:concl}

The critical properties of the synchronization transitions among
replicas of chaotic and stochastic spatially extended systems have
been numerically estimated both for diffusively coupled
and for power-law interacting systems.  In particular, 
we focus on the differences between transitions dominated by 
linear and non-linear mechanisms.

 For nearest-neighbour coupling, our analysis confirm 
previous findings indicating that the transitions are
always continuous, while two distinct universality 
classes characterize the transition depending on the nature 
of the local dynamics. For continuous (resp. discontinuous)
local maps the critical properties correspond to those of
Multiplicative Noise (resp. Directed Percolation) non-equilibrium
phase transitions.

 The introduction of a power-law coupling modifies the critical properties
of the STs, in particular for (quasi-) discontinuous maps all the studied
transitions can be gathered in a unique family of universality classes
termed Anomalous Directed Percolation. The analysis of the stochastic models
reinforce the analogy between epidemic spreading mediated by unrestricted
L{\'e}vy flights and the examined STs for a twofold reason: on one side
the numerically evaluated indexes are almost identical to the ones
found in Ref.~\cite{HH98} for a lattice model reproducing anomalous DP,
on the other hand an effective Langevin equation has been derived 
coinciding with that proposed for ADP. It is worth stressing that it 
is highly nontrivial that the deterministic systems here investigated exhibit scaling
properties in quantitative agreement with those found for stochastic models, like the lattice
model studied in \cite{HH98} and the RM-model.  Moreover, the accuracy achieved in the
investigation of the STs for the the stochastic RM-model suggests that these models can represent a valid alternative 
to the use of lattice dynamics for the investigations of non-equilibrium 
phase transitions. 

The study of smooth continuous maps 
has revealed a behaviour of the critical exponents similar to the one observed
for ADP, namely the indexes vary with continuity with the power
exponent $\sigma$, albeit their values are different
from those found for ADP. Moreover, the estimated exponents do not correspond
to any known universality class and this represents a challenge
for future theoretical investigations. In particular, the natural
candidate to explore is the field equation 
(\ref{eq:langevin1}) with noise amplitude $g(n) \propto n$
in order to understand if an 
{\it Anomalous Multiplicative Noise} class could be defined.

\acknowledgments
We thank F. Ginelli for continuous and fruitful exchanges.

\appendix

\section{Field description for the RM model}
The aim of this appendix is to provide a self-contained heuristic
derivation of the field equation associated to the ST for spatially
extended systems with power-law decaying interactions, where the 
transition is controlled by finite amplitude effects. 
In particular, we focus on the Random Multiplier model
(\ref{RMmodel1}) for $\Delta =0$, which closely reproduces
the ST for discontinuous (or quasi-discontinuous) coupled maps. 
In our derivation we follow
Ref.~\cite{GLPT03} where, by introducing a suitable spatio-temporal
coarse-graining, it has been shown that the nearest-neighbor version
of the model can be ``effectively'' described by the RFT 
associated to ordinary DP. In the following, we show
that the power law coupled version of the model can be reduced to
Eq.~(\ref{eq:langevin}), which was proposed in Ref.~\cite{HH98}
to describe ADP in epidemic spreading processes mediated by L{\'e}vy flights.

For the fully coupled case, the RM model
(\ref{RMmodel1}) can be rewritten as follows :
\begin{equation}
{v}_i(t) =  (1+\nabla^2_{\epsilon}+ \nabla^\sigma_{\epsilon})w_i(t)
\label{eq:modelloriscritto}
\end{equation}
where $v_i$ plays the role of $\tilde{w}_i$, and 
$\nabla^2_{\epsilon}$ is the discretized Laplacian operator:
\begin{equation}
\nabla^2_{\varepsilon}w_i(t) = \frac{\varepsilon}{2}w_{i+1}(t) 
+ \frac{\varepsilon}{2}w_{i-1}(t) - \varepsilon w_{i}(t) \quad,
\label{nablaeps}
\end{equation}
while $\nabla^\sigma_{\epsilon}$ represents a discretized fractional
derivative, which corresponds to the most relevant term in the small
momentum expansion of the the following discretized convolution sum
\begin{equation}
\nabla^\sigma_{\epsilon} w_i(t) \sim
\frac{\epsilon}{\eta^\prime(\sigma)}
\sum_{m=2}^{M} \frac{w_{i-m}+w_{i+m}}{m^{\sigma+1}}\,
\quad.
\label{nablasigma}
\end{equation}
Notice that the sum already contains a short distance cutoff that
should be anyway considered to have a meaningful definition of
fractional derivatives \cite{janssen,HH98}.  The stochastic variable
$w_i(t) \in [0:1]$ evolves according to (\ref{RMmodel1}), and the
positive parameter $\epsilon$ represents the amplitude of the spatial
coupling. The constant $\eta^\prime(\sigma)=2 \sum_{m=2}^{M}
(m)^{-\sigma-1}$ is a normalization factor. Periodic boundary
conditions are imposed.

Let us formally rewrite Eq. (\ref{RMmodel1}) as
\begin{equation}
w_i(t+1)= 2a v_i(t) -a^2 v_i^2(t) + g(v) \xi^\prime_i (t)
\label{eq:newevolution}
\end{equation}
where the term $\xi^\prime_i$ represents a zero-average
$\delta$-correlated noise term with unitary variance.  In order to
recognize that the above expression recovers the original model it is
enough to notice that
\begin{equation}
\xi^\prime_i (t) = \frac{1}{g(v)}[\xi_v(i,t)-\langle \xi_v(i,t) \rangle] \quad,
\end{equation}
$\xi_v$ being the dicotomic noise term
\begin{equation}
\xi_v(i, t) =
\left\{
\begin{array}{ll}
1, & \mbox{w.p.} \quad p=a{v}_i(t) \\
a{v}_i(t), & \mbox{w.p.} \quad 1-p
\end{array}
\right. \quad.
\label{2noise}
\end{equation}
whose average $\langle \xi_v \rangle$ and variance $g^2(v)$ have the
following expressions: $\langle \xi_v\rangle \!=\! (2a\!-\!1) {v}\!
-\! a^2 v^2$ and $g^2(v)\!=\! av\!-\! 3
a^2v^2\!+\!3a^3v^3\!-\!a^4v^4$ (for details see ~\cite{GLPT03}).

We can now introduce a coarse-grained variable $n(x,t)= {\bar w_i(t)}$
(where the bar denotes an average over a suitable space-time cell), in
terms of which (\ref{eq:modelloriscritto}) can be written as
$${\bar v_i(t)} = n(x,t) + \frac{\epsilon}{2} \nabla^2 n(x,t) +
\frac{c(\epsilon)}{2} \nabla^\sigma n(x,t)$$ 
where the constant $c(\varepsilon)$ takes in account the presence of
the cutoff and various normalization factors. The coarse-grained
evolution equation is then derived from (\ref{eq:newevolution}) and
reads
\begin{eqnarray}
&& \partial_t n(x,t) = (2a-1)n(x,t) + a \varepsilon \nabla^2 n(x,t) +
  a c(\varepsilon) \nabla^\sigma n(x,t) \nonumber \\ && -a^2 n^2(x,t)
  -\frac{a^2 }{4} (\varepsilon \nabla^2 n(x,t)+c(\varepsilon)
  \nabla^\sigma n(x,t) )^2 \nonumber \\ && -a^2 n(x,t) (\varepsilon
  \nabla^2 n(x,t) + c(\varepsilon) \nabla^\sigma n(x,t) ) \nonumber
  \\ && +g \left( n(x,t) + \frac{\epsilon}{2} \nabla^2 n(x,t) +
  \frac{c(\epsilon)}{2} \nabla^\sigma n(x,t) \right) \rho(x,t)
\end{eqnarray}
where the coarse-grained noise term $\rho(x,t)$ is Gaussian and
space-time $\delta$ correlated.  In proximity of the transition the
terms of order $(\nabla^2 n)^2$, $(\nabla^\sigma n)^2$, $n \nabla^2
n$, and $n \nabla^\sigma n$ can be shown to be irrelevant, and also
the terms $\sqrt{\nabla^2 n}$ and $\sqrt{\nabla^\sigma n}$ entering in
the noise amplitude $g( \dots)$.

By discarding the irrelevant terms one obtains:
\begin{eqnarray}
\partial_t n &=& \left((2a-1) + a \varepsilon \nabla^2 + a
c(\varepsilon) \nabla^\sigma\right) n(x,t) 
\nonumber
\\ &&-a^2 n^2(x,t) +
\sqrt{a n} \rho(x,t)
\end{eqnarray}
that is essentially the same Langevin equation proposed to describe anomalous DP
in \cite{janssen,HH98}.


\begin{thebibliography}{99}
\bibitem{nota0} A necessary condition to observe information propagation,
even for negative transverse Lyapunov exponent, is that the value of the slope of the
considered map should be quite large, but it is not
strictly necessary that the map is 
discontinuous~\cite{TORCINI-POLITI-GRASSBERGER,cencini_torcini_pre}. 
In this sense should be interpreted the term quasi-discontinuous.
\bibitem{nota1} It should be noticed that, only for $q=2$, the sum in
Eq.~(\ref{eq:smartcoupling}) should be slightly modified
in order to avoid an overlap with the diffusive term. In particular,
the sum should  start with $m=2$ and not $m=1$.
\bibitem{nota2} In Ref.~\cite{noi} and for the results here reported
for the Bernoulli map, we used a slightly different modified
coupling, namely
$$
\tilde{z_i}=(1-\epsilon) z_i + \frac{\epsilon}{\eta(\sigma)}
\sum_{m=1}^{M} \frac{z_{i-j_m(q)}+z_{i+j_m(q)}}{(j_m(q))^\sigma}
$$ 
with $j_m(q)=q^{m}\! -\!1$, which coincides with
~(\ref{eq:smartcoupling}) only for $q=2$. For $q\neq 2$ the above 
coupling scheme does not include a diffusive interaction. Several
tests, however, have shown that the scaling laws obtained by employing
the two coupling schemes are essentially coincident
\end{thebibliography}
\end{document}